\documentclass[a4paper,11pt]{article}
\usepackage{pos}
\usepackage{bm}
\title{Application of TMD parton showers obtained with the Parton Branching approach to Drell Yan + jets production}
\author*[a]{A. Bermudez Martinez}
\author[a]{L. I. Estevez Banos}
\author[a]{H. Jung}
\author[a]{J. Lidrych}
\author[a]{M. Mendizabal}
\author[a]{S. Taheri Monfared}
\author[a]{Q. Wang}
\author[b]{H. Yang}
\affiliation[a]{DESY, Hamburg, Germany}
\affiliation[b]{School of Physics, Peking University, Beijing, China}
\emailAdd{armando.bermudez.martinez@desy.de}
\abstract{Calculations of Drell-Yan (DY) production at next-to-leading (NLO) order have been combined with Transverse Momentum Dependent (TMD) distributions obtained with the Parton Branching (PB). The predictions show a remarkable agreement with DY measurement from E605 experiment, consistent with previous results we obtained for R209, PHENIX, CMS and ATLAS experiments. We also present predictions for Z+jet measurements showing the importance of TMD parton shower contributions to the jet multiplicity. We show that PB-TMD parton density and the corresponding PB-TMD parton shower can be combined with leading-order (LO) matrix element using the newly developed TMD merging algorithm to obtain a very good description of measurements over a wide kinematic range.}
\FullConference{%
  *** The European Physical Society Conference on High Energy Physics (EPS-HEP2021), ***\\
  *** 26-30 July 2021 ***\\
  *** Online conference, jointly organized by Universität Hamburg and the research center DESY ***
}
\begin{document}
\maketitle
\section{TMD effects in Drell-Yan production at low transverse momentum}
\begin{figure}[hbtp]
  \begin{center}
        \includegraphics[width=.45\textwidth]{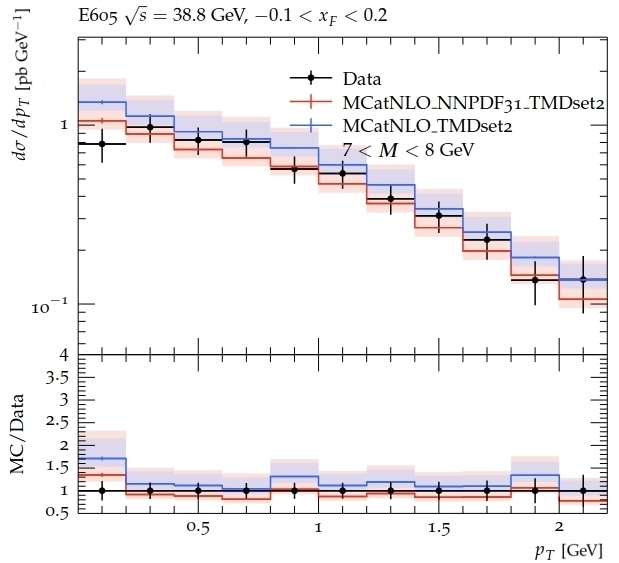}
  \caption{Transverse momentum spectrum of Drell-Yan production measured by E605~\cite{Moreno:1990sf} compared to predictions at NLO using PB-TMD density}
  \label{fig1}
  \end{center}
\end{figure}

Advanced formalisms have been developed which treat Transverse Momentum Dependent (TMD) parton density effects in inclusive observables~\cite{Angeles-Martinez:2015sea}. It was noted in~\cite{Bacchetta:2019tcu} that perturbative fixed-order calculations in collinear factorization are not able to describe the measurements of Drell-Yan (DY) transverse momentum spectra at fixed-target experiments in the region p$_\text{T}$/m$_\text{DY}\sim 1$. In~\cite{BermudezMartinez:2020tys} we have shown a first-time simultaneous description of Drell-Yan (DY) transverse momentum spectrum in a wide range of center-of-mass energy and DY mass, with comparisons to measurements from the CMS~\cite{CMS:2019raw}, NuSea~\cite{NuSea:2003qoe,Webb:2003bj}, R209~\cite{Antreasyan:1981eg}, and PHENIX~\cite{PHENIX:2018dwt} experiments. We present in Fig.~\ref{fig1} additional results for the measurement of DY production in the di-muon channel as measured by the E605 experiment~\cite{Moreno:1990sf} at a center-of-mass energy of 38.8 GeV for a DY mass in the range 7-8 GeV. A very good descrition of the data was achieved by the PB-TMD + NLO calculation (MCatNLO\_TMDset2), supporting our previous findings in~\cite{BermudezMartinez:2020tys}. A similar description is achieved when the integrated TMD distribution used in the matrix element calculation~\cite{Martinez:2018jxt} is replaced by NNPDF3.0~\cite{NNPDF:2014otw} (MCatNLO\_NNPDF31\_TMDset2), while keeping the TMD density for the generation of the transverse momentum of the incoming partons.
The calculations employed the parton branching (PB) formulation of TMD evolution set out in~\cite{Hautmann:2017fcj}. In this report we go a step further and show results for both inclusive and exclusive multi-jet observables which incorporate simultaneously the contribution of different jet multiplicities in the final state and the contribution of initial-state TMD evolution. We also show results using a new merging strategy based on the PB method to treat the evolution of the TMD parton densities and TMD parton showers and on the MLM method~\cite{Mangano:2001xp,Alwall:2007fs} to enforce exclusivity of different jet multiplicities. 
\section{PB method} 
The PB method~\cite{Hautmann:2017fcj} uses the unitarity picture of parton evolution~\cite{Webber:1986mc,eswbook}, commonly used in showering algorithms, for both collinear and TMD parton distributions. Soft gluon emission and transverse momentum recoils are treated by introducing the soft-gluon resolution scale $z_M$~\cite{Hautmann:2017xtx} to separate 
resolvable and non-resolvable branchings. The  evolution without resolvable branching from 
one scale $\mu_0$ to another scale $\mu$ is expressed through Sudakov form factors 
\begin{small}
\begin{equation}
\Delta_a(\mu^2, \mu_0^2)  = \exp\left[-\sum_b \int_{\mu_0^2}^{\mu^2}\frac{\textrm{d}\mu^{\prime 2}}{\mu^{\prime 2}} \int_0^{z_M}\textrm{d}z \  
\ z \ P_{ba}^{(R)}\left(z,\alpha_s\right) \right]\;, 
\label{eq1}
\end{equation}
\end{small}
where $a$, $b$ are flavor indices, $z$ is the longitudinal momentum fraction, $\alpha_s$ is the strong coupling, and 
$P_{ba}^{(R)} $ are real-emission splitting functions, obtained as power series expansions in $\alpha_s$. In this 
approach the TMD evolution equations are written as 
\begin{small}
\begin{eqnarray}
&& \widetilde{A}_a\left( x, {\bm k}, \mu^2\right) = \Delta_a\left(\mu^2, \mu_0^{2}\right)\widetilde{A}_a\left( x, {\bm k}, \mu_0^2\right)+ 
 \sum_b\int \frac{\textrm{d}^2{\boldsymbol \mu}^{\prime}}{\pi {\mu}^{\prime 2}}\Theta\left(\mu^{2}-\mu^{\prime 2}\right)\Theta\left(\mu^{\prime 2}-\mu_0^{ 2}\right)
 \nonumber \\ 
 &\times& 
  \int_x^{1}\textrm{d}z \ \Theta\left(z_M (\mu^\prime) - z \right)
  {  { \Delta_a\left(\mu^2, \mu_0^2  \right)  } \over 
  { \Delta_a\left(\mu^{\prime 2}, \mu_0^2 \right) } } \ 
  P_{ab}^{(R)}\left(z,\alpha_s(b(z)^2\mu^{\prime 2})\right)\widetilde{A}_b\left( \frac{x}{z},  {\bm k} + a(z){\boldsymbol \mu}^\prime, \mu^{\prime 2}\right) \; , 
\label{eq2}
\end{eqnarray} 
\end{small}
where $\widetilde{A}_a\left( x, {\bm k}, \mu^2\right)= x A_a\left( x, {\bm k}, \mu^2\right)$ is the momentum-weighted 
TMD distribution of flavor $a$, carrying the longitudinal momentum  fraction $x$ of the hadron's momentum and  transverse momentum ${\bm k}$ at the evolution scale $\mu$,  $\mu_0$ is the initial evolution scale, and  $ \mu^\prime = \sqrt{ {\boldsymbol \mu}^{\prime 2}}$ is the momentum scale at which the branching occurs.  The functions $a(z)$ and $b(z)$  in Eq.~(\ref{eq2})  specify the  ordering variable and choice of scale in the strong coupling.  If these are taken of the form prescribed by angular ordering, it is shown in~\cite{Hautmann:2019biw} that  Eq.~(\ref{eq2})  returns, once it is integrated over transverse momenta, the coherent-branching equation of~\cite{Catani:1990rr}. Numerical solutions to Eq.~(\ref{eq2}) at NLO have been used, along with  NLO calculations of Drell-Yan (DY) production in the  {\sc MadGraph5\_aMC@NLO}~\cite{Alwall:2014hca} framework, to compute  vector-boson transverse momentum spectra at LHC energies~\cite{Martinez:2019mwt} and fixed-target energies~\cite{BermudezMartinez:2020tys}, and to extract TMD densities from fits~\cite{Martinez:2018jxt} to precision deep inelastic scattering (DIS) HERA data~\cite{H1:2015ubc}. Once the evolution scale is specified in terms of kinematic variables, the PB method allows the explicit calculation of the kinematics at every branching vertex. In this approach, once the TMD distribution $ \widetilde{A}_a\left( x, {\bm k}, \mu^2\right)$ evaluated at the scale $\mu^{2}$ is known, the corresponding TMD parton shower can be generated by backward evolution. 
\section{TMD effects in Drell-Yan + jets production}
\begin{figure}[hbtp]
  \begin{center}
        \includegraphics[width=.45\textwidth]{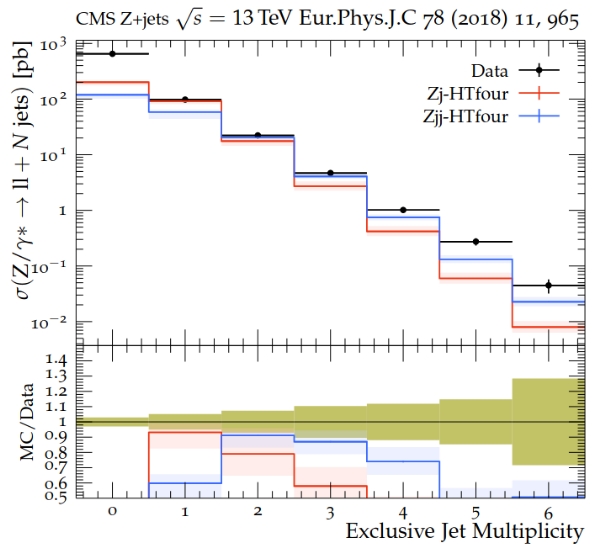}
        \includegraphics[width=.45\textwidth]{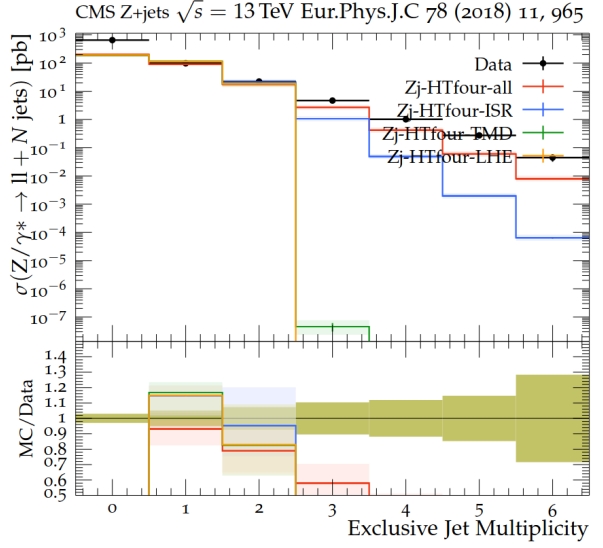}
  \caption{Exclusive jet multiplicity in the production of a Z boson in association with jets. Experimental measurement by ATLAS~\cite{ATLAS:2017sag} at 13 TeV is compared with results from (left) Z+1 and Z+2 jets TMD NLO calculations, and (right) results from Z+1 jet TMD NLO computations with and without TMD evolution and parton showers. }
  \label{fig2}
  \end{center}
\end{figure}

\begin{figure}[hbtp]
  \begin{center}
        \includegraphics[width=.45\textwidth]{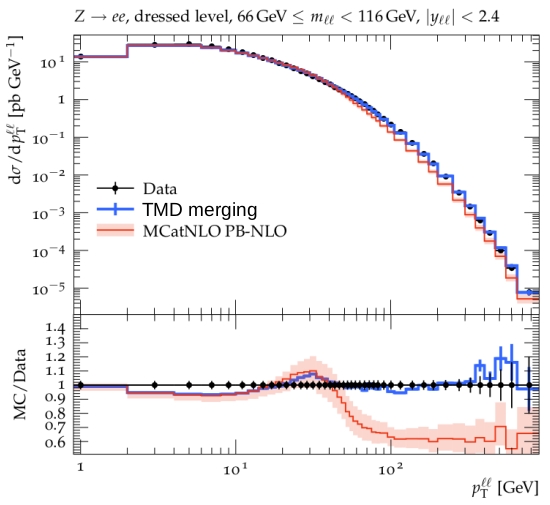}
        \includegraphics[width=.45\textwidth]{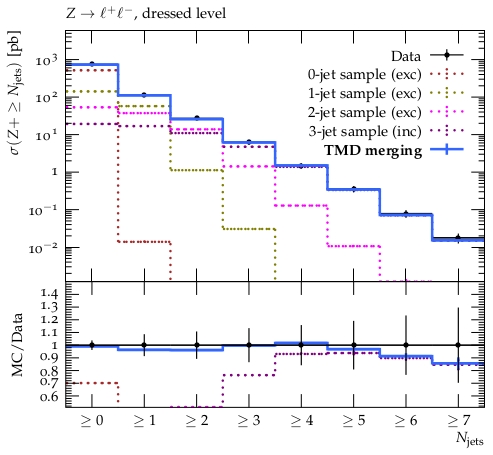}
  \caption{Left: transverse momentum p$_\text{T}$ spectrum of DY lepton pairs. Experimental measurement by ATLAS~\cite{Aad:2015auj} at 8 TeV is compared with the TMD merging calculation~\cite{Martinez:2021chk} and the NLO calculation of~\cite{Martinez:2019mwt}. The uncertainty band on the NLO calculation represents the TMD and scale uncertainties as discussed. Right: jet multiplicity in the production of a Z boson in association with jets. Experimental measurement by ATLAS~\cite{ATLAS:2017sag} at 13 TeV is compared with results from the TMD merging calculation. Contributions from the different jet samples are also shown where jet multiplicities are obtained in exclusive (exc) mode except for the highest multiplicity which is calculated in inclusive (inc) mode.}
  \label{fig3}
  \end{center}
\end{figure}

While advanced formalisms have been developed which treat TMD effects in inclusive observables~\cite{Angeles-Martinez:2015sea}, TMD implications on the exclusive structure of final states with high jet multiplicity are not very much explored. Although we have achieved a very good description of the low and intermediate DY transverse momentum spectra in a wide range of DY masses, we took a step further and studied TMD evolution effects in DY + jets production~\cite{Martinez:2021chk,Hautmann:daoste}. We studied the contribution from the TMD evolution to jet production using the backward formulation of the PB-TMD evolution equation implemented in the CASCADE generator~\cite{Baranov:2021uol}. Figure~\ref{fig2} shows the PB-TMD + NLO calculation of the exclusive jet multiplicity in Z boson production, including parton showers. Two calculations are shown in Fig.~\ref{fig2} (left), Z + 1jet at NLO (red) and also Z + 2jets at NLO (blue). The factorization and renormalization scales were set to H$_\text{T}$/4, where H$_\text{T}$ is the scalar sum of the transverse momentum of the particles in the final state. 
Although there is an important contribution from TMD evolution and parton showers, the inclusion of higher-order fixed-order calculations is crucial for a description of jet production. In order to study the contribution to jet production from the TMD initial-state shower, and also from the final state radiation we show in Fig.~\ref{fig2} (right) these separate contributions for Z + 1 jet computation at NLO accuracy. In yellow, the matrix element calculation is shown and in green the same calculation includes the TMD contribution. In addition, the blue curve includes initial-state radiation providing exclusive jet production beyond the multiplicity given by the matrix element, while in red the final result including final-state radiation as provided by the PYTHIA6 generator~\cite{Sjostrand:2006za}. The contribution from initial-state radiation provided by the TMD evolution results crucial for the description of jet production, while the final state radiation provides an additional important correction for the jet cone size utilized in the measurements.
In order to combined the TMD evolution contributions, including its exclusive manifestation in the form of an initial-state shower, and the necessary higher-order corrections we developed a new TMD merging algorithm~\cite{Martinez:2021chk} at LO level expanding on the MLM matching prescription~\cite{Mangano:2001xp,Alwall:2007fs}. Similar constructions based on other approaches are possible. Compared to standard collinear merging frameworks the TMD merging algorithm as distinctive features: i) for any n-jet parton level event, initial-state transverse momenta are generated according to the TMD distributions obtained as solutions of the PB evolution equations~\cite{Hautmann:2017fcj}, but rejecting, owing to Sudakov suppression, events where the transverse momentum provided by the TMD evolution is larger than the minimum energy scale in the hard process; ii) initial state partons in the process are showered using the backward space-like shower evolution given by the PB-TMD equations while final state partons are showered using standard time-like showers; iii) a merging prescription, such as MLM~\cite{Mangano:2001xp,Alwall:2007fs}, is applied between the showered event and the event generated in i) including the k$_\text{T}$ boost.
In Fig.~\ref{fig3} we proceed to compute the transverse momentum spectrum of DY lepton pairs from $Z$-boson decay as well as the the jet multiplicity in DY events. The results are compared with ATLAS measurements~\cite{Aad:2015auj,ATLAS:2017sag}.  
The LO result is normalized to the next-to-NLO (NNLO) DY cross section. The contributions from the different jet multiplicities to the final prediction are shown separately.        
The merged prediction is found to provide a good description of the data throughout the whole Z boson p$_\text{T}$ spectrum, while consistently keeping the good level of agreement at low and intermediate Z p$_\text{T}$ compared to our result in~\cite{Martinez:2019mwt}. For the case of Fig. 3 (right) we observe good ageement even for multiplicities much larger than the maximum number of jets (three) for which the exact LO matrix element calculation is performed. This highlights the potential benefit of the TMD evolution in the description of hard and non-collinear emissions, compared to standard collinear evolution.
\section{Conclusions} 
We have presented new results for the DY p$_\text{T}$ spectrum at low center-of-mass energy and low DY mass that confirm the conclusions in~\cite{BermudezMartinez:2020tys}. We have also studied the contributions from TMD evolution to jet production and found that it is crucial in the description of jet observables. We have also shown results using the new TMD merging method to analyze jet final states in high-energy hadronic collisions. While TMD effects have mostly been studied so far in the low-p$_\text{T}$ inclusive spectra, our approach allows to initiate investigations of possible TMD effects at the level of exclusive jet observables and in the domain of the highest p$_\text{T}$ processes, where the impact on searches for signals of new physics may be largest. An excellent description of the DY spectrum in a wide range of p$_\text{T}$, as well as of the jet multiplicity even much beyond the reach of the corresponding fixed-order caclulation were found.

\end{document}